\documentclass[usenatbib,usegraphicx]{mn2e}
\usepackage{amsfonts}
\usepackage{amsmath}
\usepackage{amssymb}
\usepackage{comment}
\usepackage[usenames,dvipsnames]{color}

\title[Galaxy mergers at $z>1$]
    {Galaxy merger histories and the role of merging in driving star formation at $z>1$}
\author[S. Kaviraj et al.]
{S. Kaviraj,$^{1}$\thanks{s.kaviraj@herts.ac.uk} J.
Devriendt,$^{2}$ Y. Dubois,$^{3,4}$ A. Slyz,$^{2}$ C.
Welker,$^{3,4}$ C. Pichon,$^{3,4}$ \newauthor S. Peirani$^{3,4}$
and D. Le Borgne$^{3,4}$\\
$^{1}$Centre for Astrophysics Research, University of
Hertfordshire, College Lane, Hatfield, Herts, AL10 9AB, UK\\
$^{2}$Dept of Physics, University of Oxford, Keble Road, Oxford
OX1 3RH UK\\
$^{3}$Sorbonne Universit\'es, UPMC Univ Paris 06, UMR 7095, Institut d'Astrophysique de Paris, F-75005 Paris, France\\
$^{4}$CNRS, UMR 7095, Institut d'Astrophysique de Paris, 98 bis
Boulevard Arago, F-75014 Paris, France}

\begin{document}

\maketitle

\def \aj {AJ}
\def \mnras {MNRAS}
\def \pasp {PASP}
\def \apj {ApJ}
\def \apjs {ApJS}
\def \apjl {ApJL}
\def \aap {A\&A}
\def \nat {Nature}
\def \araa {ARAA}
\def \iaucirc {IAUC}
\def \aaps {A\&A Suppl.}
\def \qjras {QJRAS}
\def \na {New Astronomy}
\def \aapr {A\&ARv}
\def\lesssim{\mathrel{\hbox{\rlap{\hbox{\lower4pt\hbox{$\sim$}}}\hbox{$<$}}}}
\def\gtrsim{\mathrel{\hbox{\rlap{\hbox{\lower4pt\hbox{$\sim$}}}\hbox{$>$}}}}


\begin{abstract}
We use Horizon-AGN, a hydrodynamical cosmological simulation, to
explore the role of mergers in the evolution of massive (M$_*>
10^{10}$ M$_{\odot}$) galaxies around the epoch of peak cosmic
star formation ($1<z<4$). The fraction of massive galaxies in
major mergers (mass ratio $R<4:1$) is around 3\%, a factor of
$\sim$2.5 lower than minor mergers ($4:1<R <10:1$) at these
epochs, with no trend with redshift. At $z\sim1$, around a third
of massive galaxies have undergone a major merger, while all such
systems have undergone either a major or minor merger. While
almost all major mergers at $z>3$ are `blue' (i.e. have
significant associated star formation), the proportion of `red'
mergers increases rapidly at $z<2$, with most merging systems at
$z\sim1.5$ producing remnants that are red in rest-frame
UV-optical colours. The star formation enhancement during major
mergers is mild ($\sim$20-40\%) which, together with the low
incidence of such events, implies that this process is not a
significant driver of early stellar mass growth. Mergers ($R <
10:1$) host around a quarter of the total star formation budget in
this redshift range, with major mergers hosting around two-thirds
of this contribution. Notwithstanding their central importance to
the standard $\Lambda$CDM paradigm, mergers are minority players
in driving star formation at the epochs where the bulk of today's
stellar mass was formed.
\end{abstract}




\section{Introduction}

The cosmic star formation history indicates that the bulk of the
stars in today's massive galaxies formed at $z \sim 2$
\citep[e.g.][]{Madau1998,Hopkins2006}. While the nearby Universe
underpins much of our current understanding of galaxy evolution, a
convergence of new multi-wavelength surveys
\citep[e.g.][]{Windhorst2011,Grogin2011} and high-resolution
hydrodynamical simulations
\citep[e.g.][]{Devriendt2010,Duboisetal14,Schaye2014,Khandai2014,Vogelsberger2014}
is revolutionizing our understanding of the $z>1$ Universe. The
processes that drive galaxy growth at these early epochs are still
the subject of much debate. The role of mergers (major mergers in
particular) at high redshift remains poorly understood. While
mergers are capable of inducing strong star formation, black-hole
growth and morphological transformations
\citep[e.g.][]{Springel2005a}, both theory
\citep[e.g.][]{Keres2009,Dekel2009a} and observation (e.g. Genzel
et al. 2011; Kaviraj et al. 2013a, K13a hereafter) have recently
suggested a lesser (perhaps insignificant) role for this process
in driving stellar mass growth in the early Universe.

Spectroscopic studies of star-forming galaxies around $z\sim2$
have demonstrated high fractions of systems that are not in
mergers but show kinematic morphologies indicative of turbulent
discs
\citep[e.g.][]{ForsterSchreiber2006,Genzel2008,Shapiro2008,Law2009,Mancini2011}.
Since galaxy morphology carries an imprint of the mechanisms that
drive star formation \citep[e.g.][]{Lintott2011}, imaging studies,
using e.g. high-resolution instruments like the \emph{Hubble Space
Telescope}, have offered a complementary route to probing the
processes that dominate stellar mass growth at $z>1$. In broad
agreement with the spectroscopic literature, a possible dominance
of non-merging galaxies within the early star-forming population
has been suggested by such work (e.g. Lotz et al. 2006;
F{\"o}rster Schreiber et al. 2011; Law et al. 2012; Kaviraj et al.
2013a). In particular, Kaviraj et al. (2013a; K13a hereafter) have
probed the overall significance of (major) merger-driven star
formation in the early Universe, by estimating the fractional
contribution of visually-classified mergers to the star formation
budget at $z\sim2$. Their work estimates that less than a third of
the budget at this redshift is hosted by systems that are in major
mergers.

While observational studies are transforming our understanding of
the role of mergers at $z>1$, the empirical literature unavoidably
uses a heterogeneous set of methodologies and is often restricted
to relatively small galaxy samples, with each study typically
probing a relatively narrow range in redshift. A complementary
approach to exploring the role of merging is to use a theoretical
model that is well-calibrated to the observed Universe at these
epochs. In this study, we exploit Horizon-AGN, a hydrodynamical
cosmological simulation that reproduces the mass functions and
rest-frame UV-optical colours of massive galaxies at $z>1$, to
probe the impact of merging on massive (M$_*> 10^{10}$
M$_{\odot}$) galaxies at $1<z<4$. Our principal objectives are to
perform a statistical study of major/minor merger histories, the
enhancement of star formation in major mergers and the proportion
of the star formation budget that is attributable to mergers
around the epoch of peak star formation.

This Letter is organized as follows. In Section 2, we describe the
simulation that underpins our study. In Section 3, we study the
fractions of galaxies in major and minor mergers at $1<z<4$, the
cumulative merger histories of massive galaxies at $z\sim1$ and
the star formation activity in mergers across this epoch. In
Section 4, we study the enhancement of star formation induced by
the major-merger process and calculate the fraction of the star
formation budget that is hosted by merging systems at these
redshifts. We summarize our findings in Section 5.


\section{The Horizon-AGN simulation}
We begin by summarizing the main properties of the Horizon-AGN
simulation - we refer readers to \citet{Duboisetal14} for further
details. The simulation employs the adaptive mesh refinement
Eulerian hydrodynamics code, RAMSES \citep{Teyssier2002}. The size
of the simulated volume is 100 $h^{-1}\, \rm Mpc$ comoving,
containing $1024^3$ dark matter particles with initial conditions
that correspond to a standard $\Lambda$CDM cosmology with
Planck-like values \citep{Planck}. The initial $1024^3$ uniform
grid is refined with a quasi Lagrangian criterion when 8 times the
initial total matter resolution is reached in a cell, down to a
minimum cell size of $1 \, \rm kpc$ in proper units.

Metal-dependent radiative cooling is implemented following
\citet{Sutherland1993} and a uniform UV background is switched on
at $z = 10$, following \citet{Haardt1996}. The standard
Schmidt-Kennicutt law \citep[e.g.][]{Kennicutt1998} is employed to
produce star particles with a 2\% efficiency, when the gas density
reaches a critical density of 0.1 H cm$^3$. Based
on~\cite{Dubois2008}, mass loss from massive stars occurs via
stellar winds and Type II and Type Ia supernovae, which disperse
gas and metals into the ambient medium (Kimm et al, in prep.).
Seed black holes (BHs) with a mass of 10$^5$ M$_{\odot}$ are
assumed to form in regions of high gas density and the growth of
the BH is tracked self consistently based on a modified Bondi
accretion rate. When gas accretes on to BHs, we assume that a
central BH impacts the ambient gas in two possible ways, depending
on the accretion rate. For a high accretion rate (Eddington ratio
$>0.01$), 1.5 per cent of the accretion energy is injected as
thermal energy (a quasar-like feedback mode), while jets are
employed for low accretion rates (Eddington ratios $<0.01$) with a
10 per cent efficiency \citep{Dubois2012}. Due to the presence of
AGN feedback, the mass functions of massive galaxies and the
rest-frame UV-optical colours of observed galaxies around the
epoch of peak cosmic star formation are well reproduced by our
model \citep{Kimm2012, Duboisetal14}. Given the high sensitivity
of the rest-frame UV wavelengths to star formation
\citep[e.g.][]{Martin2005,Kaviraj2007c}, the agreement with
UV-optical colours (and mass functions) implies a good
reproduction of galaxy star formation histories in the simulation,
making it a useful tool for exploring the processes that drive
stellar mass growth at these epochs.

We build merger trees using TreeMaker \citep{Tweed2009}, with a
typical time difference between time steps of $\sim$35 Myr (the
range is between 20 and 60 Myr). Here, we explore the merger
histories of the set of $\sim1100$ Horizon-AGN galaxies at
$z\sim1.2$ that have masses greater than $10^{10}$ M$_{\odot}$.
The minimum galaxy mass probed is $\sim10^{8.5}$ M$_{\odot}$,
making the simulation complete to mergers with mass ratios of
$\sim10:1$ or less. Our subsequent analysis is restricted to this
range of mass-ratio values. In what follows, mergers are defined
as systems that merge within a timescale of $\sim$0.1 Gyrs. These
galaxies are at the final stages of a merger, with the centres of
the merger progenitors typically less than 20 kpc from each other.
Tidal features are more readily observable in images at such
separations \citep[e.g.][]{Darg2010a,Darg2010b}, making our model
results better aligned with observations of `close pairs' systems
at similar relative distances.

\begin{figure}
$\begin{array}{c}
\includegraphics[width=3.5in]{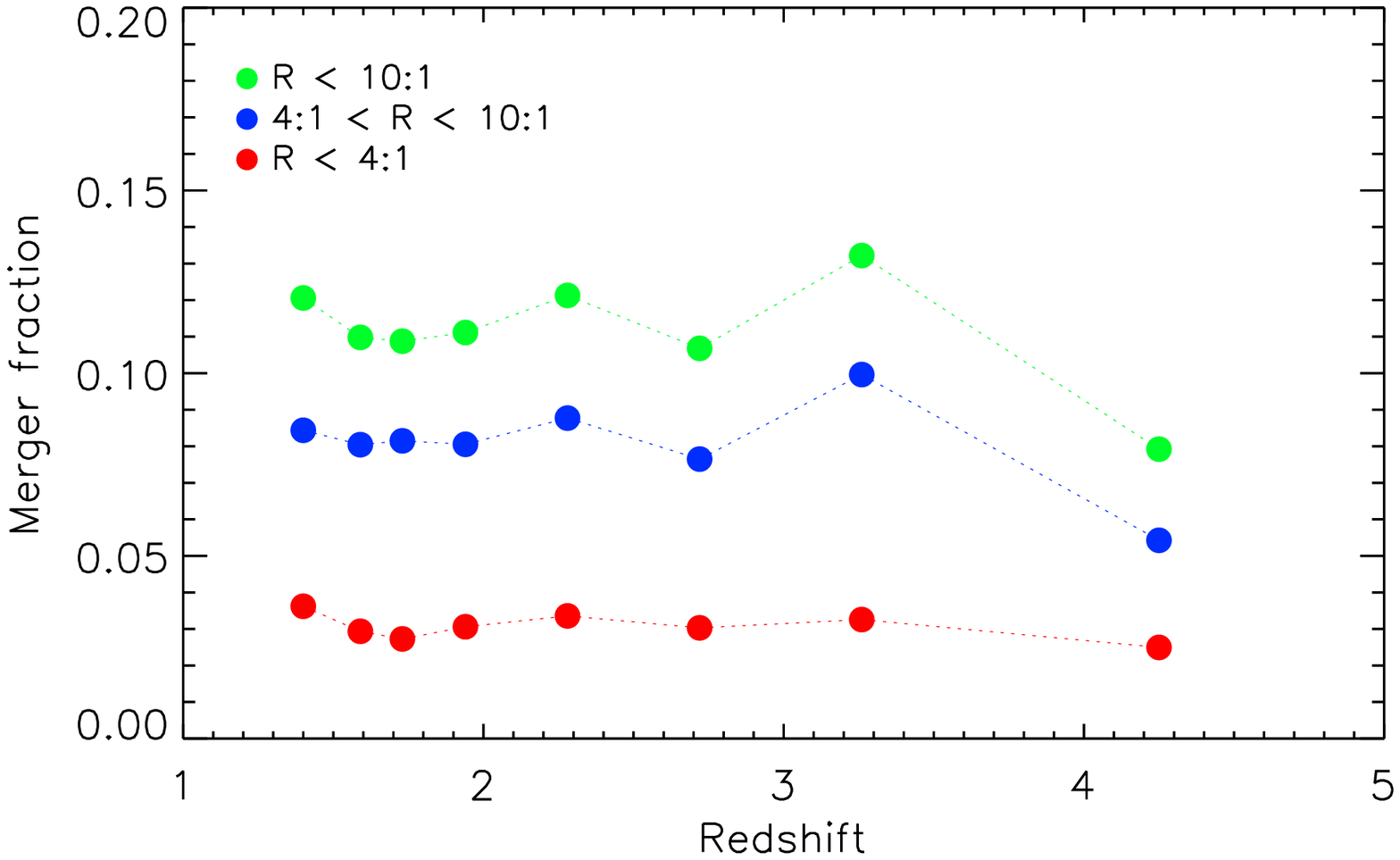}\\
\includegraphics[width=3.5in]{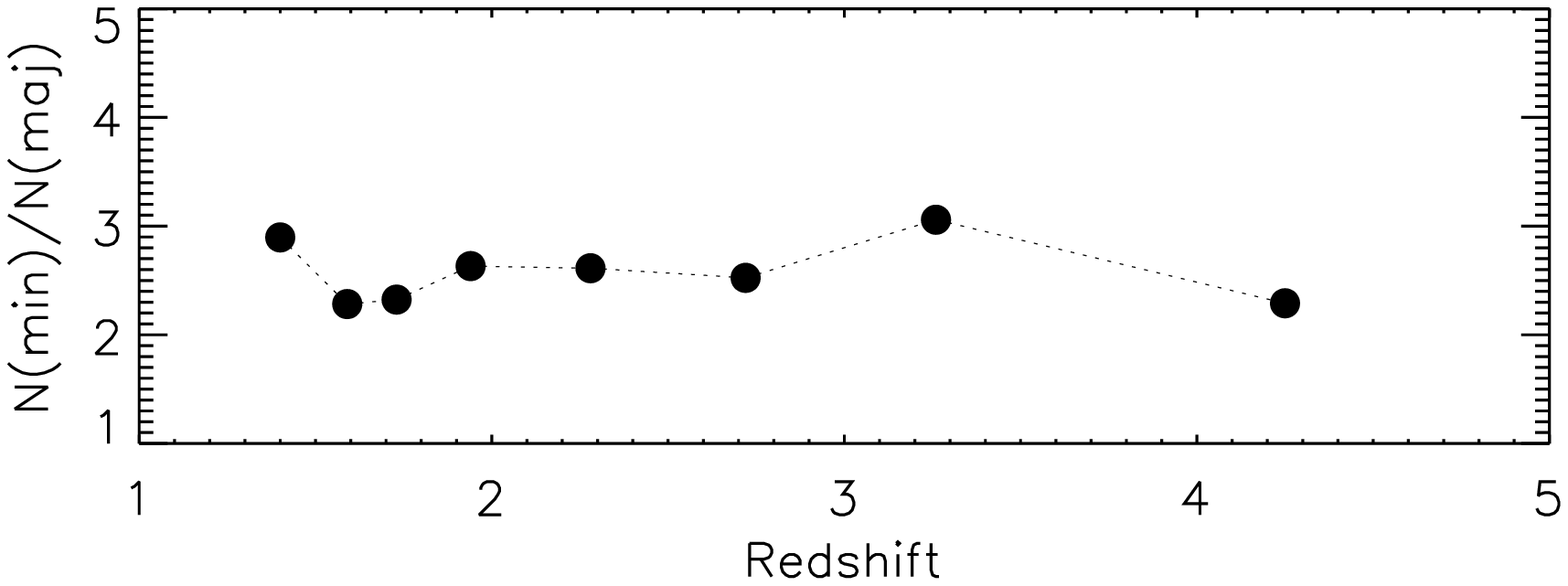}
\end{array}$
\caption{\textbf{Top:} Fractions of massive galaxies in mergers of
different mass ratios ($R$). The red line indicates major mergers
($R<4:1$), the blue line indicates minor mergers ($4:1<R<10:1$),
while the green line indicates all mergers with R $<10:1$.
\textbf{Bottom:} The ratio of minor to major mergers (i.e. the
ratio of the red and blue curves in the panel above).}
\label{fig:merger_fractions}
\end{figure}

\begin{figure}
$\begin{array}{c}
\includegraphics[width=3.5in]{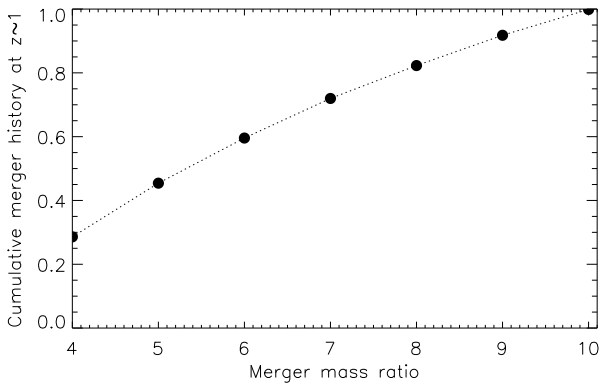}
\end{array}$
\caption{The fraction of massive galaxies at $z\sim1$ that have
had a merger with a mass ratio ($R$) less than the value on the
x-axis. For example, at $z\sim1$, 30\% of galaxies have had a
major merger ($R<4:1$), while all massive galaxies have had either
a major or minor merger with ($R<10:1$).}
\label{fig:merger_histories}
\end{figure}


\section{Merger histories of massive galaxies}
We begin by exploring the merger histories of massive galaxies
around the epoch of peak cosmic star formation. In the top panel
of Fig. \ref{fig:merger_fractions}, we present the fractions of
galaxies in mergers of various mass ratios ($R$), where $R$ is
defined as the mass of the larger progenitor divided by the mass
of its smaller counterpart. The fractions of merging systems with
$R<4:1$ (`major' mergers), $4:1<R<10:1$ (`minor' mergers) and
$R<10:1$ (major + and minor mergers) are around 3\%, 8\% and 11\%
respectively, with no trend with redshift. In the redshift range
$1<z<4$, the number of minor mergers is around a factor of 2.5
higher than their major counterparts (bottom panel of this
figure). In Fig. \ref{fig:merger_histories}, we present a
cumulative view of the average merger history of massive galaxies
at $z\sim1$. We show the fraction of galaxies that have had a
merger with a mass ratio less than the value on the x-axis. Thus,
by $z\sim1$, 30\% of galaxies have had a major merger, while all
massive galaxies have had either a major or minor merger with
$R<10:1$. Note that we do not consider mass ratios greater than
$10:1$ because, as noted above, the simulation is incomplete
beyond these mass ratios across our redshift range of interests.

\begin{figure}
$\begin{array}{c}
\includegraphics[width=3.5in]{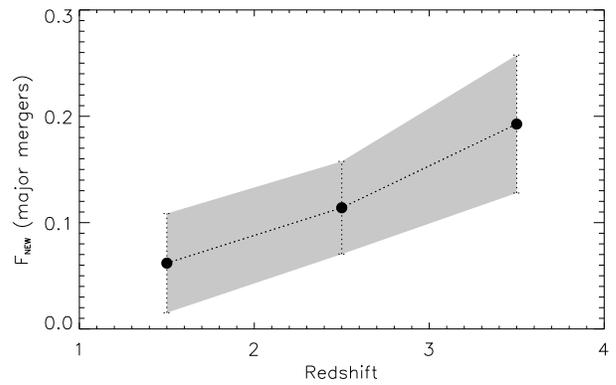}
\end{array}$
\caption{The mass fraction ($F_{new}$) of young (age $< 0.1$ Gyr)
stars in major-merger remnants. The points indicate median values,
while the width of the shaded region indicates the standard
deviation in the values of $F_{new}$ at that redshift.}
\label{fig:dry_wet_mergers}
\end{figure}

The major merger fractions in our model and their lack of redshift
evolution are consistent with both $\Lambda$CDM values from the
Millenium simulation \citep[e.g.][]{Bertone2009} and empirical
merger fractions \citep[e.g.][]{Man2012}. The cumulative merger
history estimated by observational studies suggests that a massive
galaxy experiences $\sim$1.1 major mergers in the redshift range
$0<z<3$ (Man et al. 2012, see also Bluck et al. 2012). Assuming
the merger rate does not evolve strongly with redshift, as
suggested by this study and others
\citep[e.g.][]{Lotz2011,Man2012}, this indicates that around a
third of massive galaxies will have experienced a major merger by
$z \sim1$, in agreement with the cumulative merger histories
presented in Fig. \ref{fig:merger_histories}. Finally, the ratio
of minor to major mergers ($\sim \times 2.5$) in our model is also
consistent with recent observational work
\citep{Lotz2011,Bluck2012}.

It is worth noting here that the observational literature on
mergers is still developing at these epochs, and that the
dispersion in current observational estimates of the merger
fraction can be significant, with different studies reporting
values that are discrepant by several factors \citep[see e.g. the
compilation of results in figure 12 of][]{Conselice2014}. The
discrepancies are likely to be driven by different selection
techniques, low number statistics in the high-redshift fields and
cosmic variance \citep[e.g.][]{Lotz2011}. The large current
dispersion in the observational data makes a theoretical study
such as this one useful, both for providing a quantitative picture
of galaxy merging at $z>1$, and for offering theoretical estimates
that can be better tested as the observational literature matures.

In Fig. \ref{fig:dry_wet_mergers} we study the star formation
activity in major mergers. Since the literature often invokes red
mergers (i.e. those where the fraction of new stars formed is
close to zero) to explain massive-galaxy evolution, it is useful
to explore the star formation that is associated with mergers at
these epochs. We define the fraction of `new' stars as the
fraction of stellar mass formed within the last 0.1 Gyr
($F_{new}$), which is close to typical dynamical timescales at
these epochs \citep[e.g.][]{Ceverino2010}. The points in this
figure indicate median values, while the width of the shaded
region indicates the standard deviation in the distribution of
$F_{new}$ values at a given redshift.

We cast our results in terms of an empirical boundary between
`red' and `blue' systems, estimated via recent observational work
at these epochs. Using star formation histories constructed via
spectral energy distribution (SED) fitting (Kaviraj et al. 2013b,
K13b hereafter) have recently studied the rest-frame UV-optical
colours of massive galaxies at $1<z<3$. The high sensitivity of
the UV wavelengths to star formation means that even small mass
fractions of young stars (of the order of a percent or less) will
drive galaxies into the UV-optical blue cloud
\citep[e.g.][]{Yi2005,Kaviraj2007a}, making the UV-optical colour
space particularly effective at separating star-forming and
quiescent systems. Taking the boundary between red and blue
galaxies to be $(NUV-V)\sim3$ (see figure 4 in K13b), the
SED-fitted star formation histories in K13b indicate that galaxies
that are blue in UV-optical colours typically have
$F_{new}$$\gtrsim0.1$. We therefore take $F_{new}=0.1$ as the
boundary between red and blue systems in our subsequent analysis.
Based on the blue vs red threshold assumed above, we find that,
while almost all major mergers at $z>3$ are blue, the proportion
of red mergers increases rapidly at $z<2$. At $z\sim1.5$, most
(but not all) major mergers are red in UV-optical colours (i.e.
produce very little star formation).


\section{Are major mergers significant drivers of star formation?}
The overall role of the major-merger process in driving stellar
mass growth has been an important question in the recent
literature (e.g. K13a; Rodighiero et al. 2011). The overall
significance of major-merger driven star formation depends both on
the frequency of such mergers and on the enhancement of star
formation that is induced when a major merger event takes place.
Thus, if the enhancement of star formation is high during the
merger, then major-merger-driven star formation can be a
significant contributor to the star formation budget, even if the
merger fraction itself is relatively low (as has been shown to be
the case in the previous section).

\begin{figure}
$\begin{array}{c}
\includegraphics[width=3.5in]{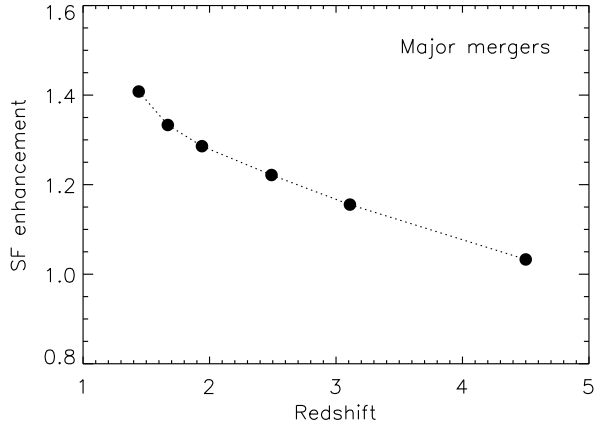}
\end{array}$
\caption{Star formation enhancement in systems that are major
mergers. The enhancement is defined as the ratio of the total star
formation activity in mergers to that in non-merging systems at a
given redshift.} \label{fig:sf_budget1}
\end{figure}

\begin{figure}
$\begin{array}{c}
\includegraphics[width=3.5in]{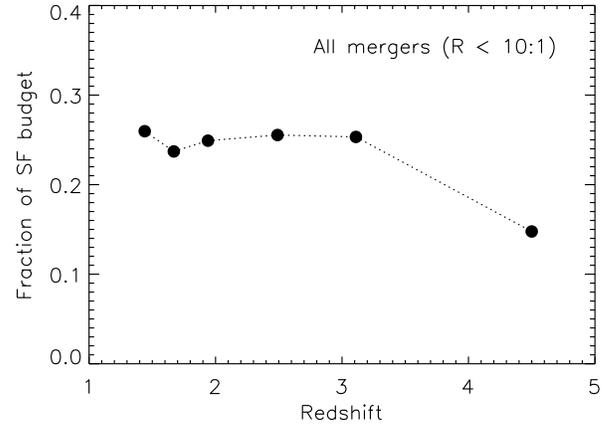}
\end{array}$
\caption{The fraction of the total star formation budget that is
hosted by systems that are in mergers with $R<10:1$. Major mergers
host around two-thirds of the contribution, without much trend
with redshift.} \label{fig:sf_budget2}
\end{figure}

In Fig. \ref{fig:sf_budget1} we present the enhancement of star
formation during major mergers. The enhancement is defined as the
average $F_{new}$ in merging systems divided by the average
$F_{new}$ in the population of galaxies that are not undergoing a
major merger. We find that the enhancement is relatively mild -
around 20-40\% in the redshift range $1<z<3$. This is smaller than
suggested by K13a who estimated an enhancement of a factor of 2,
but consistent with a forthcoming study that has repeated this
analysis on much larger sample of galaxies (Lofthouse et al., in
prep). Interestingly, the star formation enhancement in major
mergers shows a gradual increase with decreasing redshift. This is
likely driven by the gradual decrease in the `background' level of
star formation in normal (non-merging) galaxies, as has been noted
in recent observational work (e.g. K13a). In other words, as star
formation activity in galaxies becomes progressively more
quiescent, the impact of a major merger becomes proportionately
higher. Indeed in the local Universe, where star formation driven
by secular processes is significantly weaker than at high redshift
\citep[e.g.][]{Pannella2014}, gas-rich major mergers can enhance
star formation by orders of magnitude \citep[e.g.][]{Mihos1996},
in contrast with the situation at $z>1$.

Taken together with the relatively low frequency of major mergers
(recall that only $\sim$30\% of massive galaxies at $z\sim1$ have
had such a merger), this indicates that, overall, the major merger
process is not a significant driver of stellar mass growth at
these redshifts. In Fig.~\ref{fig:sf_budget2}, we quantify this
further by calculating the fraction of the star formation budget
that is hosted by major and minor mergers ($R<10:1$). This
fraction is around 25\% across our redshift range of interest.
Around two-thirds of this value (17\%) is in systems with $R<4:1$
(i.e. major mergers), again slightly lower than the observational
estimate of Kaviraj et al. (2013a) but consistent within the
observational uncertainties.


\section{Summary}
We have used Horizon-AGN, a hydrodynamical cosmological simulation
that reproduces the mass functions and rest-frame UV-optical
colours of massive galaxies at $z>1$, to probe the merger
histories of massive galaxies and the impact of merging (major
mergers in particular) in triggering stellar mass growth at
$1<z<4$. Our main conclusions are as follows:

\begin{itemize}

    \item The fraction of massive galaxies in major mergers ($R<4:1$) is
    around 3\%, while the fraction of systems in major or minor ($R<10:1$)
    mergers is around 11\%. The merger fractions show no trend with redshift in the range
    $1<z<4$.\\

    \item Minor mergers ($4:1<R<10:1$) are around a factor of
    2.5 more frequent than their major counterparts at $1<z<4$.\\

    \item At $z\sim1$, $\sim$30\% of massive galaxies have undergone a major merger, while all massive galaxies have undergone either a major or minor merger, i.e. a merger with $R<10:1$.\\

    \item While almost all major mergers at $z>3$ are blue (i.e. result in significant star formation), the proportion
of red mergers increases at $z<2$, with most major mergers at
$z\sim1.5$, producing remnants that are red in rest-frame
UV-optical colours.\\

    \item The enhancement of star formation in major mergers at
    these epochs is relatively mild ($\sim$20-40\%). Together with the low frequency of these events, this indicates that major mergers are
not
    significant drivers of stellar mass growth at these
    redshifts.\\

    \item (Only) a quarter of the total star formation budget is
    hosted by mergers with $R<10:1$. While they are a key feature
    of the standard hierarchical paradigm, mergers with $R<10:1$
    play a relatively insignificant role in driving stellar mass
    growth in the early Universe.

\end{itemize}


\section*{Acknowledgements}
SK acknowledges a Senior Research Fellowship from Worcester
College Oxford. Part of this analysis was performed on the DiRAC
facility jointly funded by STFC and the Large Facilities Capital
Fund of BIS. JD and AS's research is supported by funding from
Adrian Beecroft, the Oxford Martin School and the STFC. This
research has used the HPC resources of CINES (Jade supercomputer),
under the allocations 2013047012 and c2014047012 made by GENCI and
has been partially supported by grant Spin(e) ANR-13-BS05-0005 of
the French ANR. The post-processing has made use of the Horizon
and Dirac clusters. This work is part of the Horizon-UK project.


\nocite{Bluck2012,Genzel2011,Kaviraj2013a,Kaviraj2013b,Lotz2006,Rodighiero2011,ForsterSchreiber2011,Law2012}


\bibliographystyle{mn2e}
\bibliography{references}


\end{document}